\documentclass[twocolumn,amsmath,amssymb,showpacs,superscriptaddress,aps,longbibliography,10pt]{revtex4-1}

\usepackage{graphicx}
\usepackage{bm}
\usepackage{dcolumn}
\usepackage{amssymb}
\usepackage{amsmath}
\usepackage{amsfonts}
\usepackage{epsfig}
\usepackage{graphicx}
\usepackage{stackrel}
\usepackage{paralist}
\usepackage{hyperref}
\usepackage{color}

\newcommand{\refEq}[1]{Eq.~(\ref{#1})}
\newcommand{\refFig}[1]{Fig.~\ref{#1}}

\newcommand{\citeRef}[1]{Ref.~\cite{#1}}
\newcommand{\citeRefs}[1]{Refs.~\cite{#1}}

\begin{document}
\title{From Bloch Oscillations to Many Body Localization in Clean Interacting Systems}

\author{Evert P.~L.~van Nieuwenburg}
\thanks{These two authors contributed equally.}
\affiliation{Institute of Quantum Information and Matter, California Institute of Technology, Pasadena, California 91125, USA}
\author{Yuval Baum}
\thanks{These two authors contributed equally.}
\affiliation{Institute of Quantum Information and Matter, California Institute of Technology, Pasadena, California 91125, USA}
\author{Gil Refael}
\affiliation{Institute of Quantum Information and Matter, California Institute of Technology, Pasadena, California 91125, USA}

\begin{abstract}
In this work we demonstrate that non-random mechanisms that lead to single-particle localization may also lead to many-body localization, even in the absence of disorder. In particular, we consider interacting spins and fermions in the presence of a linear potential. In the non-interacting limit, these models show the well known Wannier-Stark localization. We analyze the fate of this localization in the presence of interactions. Remarkably, we find that beyond a critical value of the potential gradient, these models exhibit non-ergodic behavior as indicated by their spectral and dynamical properties. These models, therefore, constitute a new class of generic non-random models that fail to thermalize. As such, they suggest new directions for experimentally exploring and understanding the phenomena of many-body localization. We supplement our work by showing that by employing machine learning techniques, the level statistics of a system may be calculated without generating and diagonalizing the Hamiltonian, which allows a generation of large statistics.
\end{abstract}

\maketitle

\section{Introduction}

Since the phenomenon of many-body-localization (MBL) was re-postulated more than a decade ago \cite{AndersonLoc,BASKO,mirlin}, it has attracted a great deal of attention.
It provides an example of a generic quantum many-body system that cannot reach thermal equilibrium \cite{MBL_rev1,MBL_rev2,Imbrie,HuseOgan2}.
In recent years, an enormous theoretical effort was invested in understanding the nature of the MBL transition \cite{HuseOgan,PalHuse,vosk}, the dynamical \cite{dynamics1,dynamics2,dynamics3} and
entanglement \cite{mblChetan,XXZ,Ent1,Ent2} properties of these systems and their response to external probes \cite{resp,res2} and periodic driving \cite{drive1,drive3,drive4}.
Also the experimental community \cite{exp1,exp2,exp3,exp4,exp5} has found interest in this field, in particular, because these systems have the potential of storing information about
initial states for long times, and hence may implement quantum memory devices. These systems may also be useful for dynamical quantum control, as they allow the application of driving
protocols without heating the system to an infinite temperature.

A key ingredient for achieving the MBL phase is disorder (randomness). The roots of this phase lie within the phenomenon of Anderson localization \cite{AndersonLoc}, where non-interacting particles
form a localized non-ergodic phase. Questioning the fate of Anderson localization in the presence of interactions led to the discovery of the MBL phase.

In this work we ask whether randomness is indeed an essential ingredient in achieving generic non-ergodic interacting phases. Viewing MBL as a competition
between single-particle localization and interactions, one may wonder whether a localizing mechanism that does not require disorder may produce similar results. It was suggested that quasi-many-body
localization may exist in a translationally invariant quantum system such as a quantum disentangled liquid \cite{Roeck2,mblnodis1,mblnodis2,mblnodis3}, where light particles evade thermalization
(for long times) by localizing on heavy particles \cite{qdl,Kagan,Roeck1,Markus,qdl2}. Moreover, it was shown that clean $1$D systems with quasi-periodic potentials may host an MBL phase \cite{QPmbl1,QPmbl2,qp_mbl}.
While quasi periodic systems are not considered disordered, they do not respect exactly the discrete translational symmetry of the lattice either
and can not be treated in momentum space. Other proposals (\citeRef{Haa}) suggested the appearance of non-ergodic dynamics for a large portion of states belonging
to the low energy subspace of the cubic code Hamiltonian which involves eight-spin interaction terms. The model we propose in this work respects the crystal
symmetry exactly, and hence, in that regard it is a realizable and a truly discrete translational invariant model. We show that this model supports a phase
that is indistinguishable from the MBL phase based on all the standard characteristics.

A well known mechanism for localizing single particles is the Wannier-Stark effect \cite{SWE}, in which particles living on a lattice become localized in the presence of a linear potential.
We refer to this phenomenon as Bloch localization. Notice that beside lacking randomness, such systems also preserve translation-invariance as the linear potential represents a uniform force and may
be replaced by a time-dependent vector potential. One may consider Wannier-Stark effect as a particular case of dynamical localization \cite{dynamical_loc} with linear-in-time vector potential.
While no physical difference is expected between the different gauges, the thermodynamic limit in the time-dependent gauge avoids the existence of an infinite energy difference between the edges of the system. Nevertheless, we chose to work in the static gauge since in this work we are only interested in static forces and since it dramatically reduces the numerical effort. In the appendices we show that our numerical method is, as expected, indifferent to the choice of gauge.
The fate of dynamical-localization in the case of time dependent fields has been discussed
in \citeRefs{many_DL1,many_DL2}. The interplay between interactions and linear fields has been investigated in the past. It was shown that the oscillatory part of the current, i.e. Bloch oscillations (BO),
decays as the interaction strength increases \cite{rel1,rel3}. It was also shown that the presence of a uniform force changes the nature of the evolution of an initial state under the non-linear Schr\"{o}dinger equation (NLSE) as the non-linearity increases, e.g., for a large non-linearity the dynamics is localized. Yet, the ergodic properties and the generality (stability) of these phases can not be inferred from these works. The absence of BO does not necessarily signify ergodicity and the dynamics of generic interacting models can not be captured by the NLSE, which is generally valid only as a mean field description of weakly interacting Bosons \cite{nlse}. Moreover, only the evolution of low energy (near ground state) states have been considered and the stability of the above phenomenon was not analyzed.
\begin{figure*}[t]
\centering
\includegraphics[width =\linewidth]{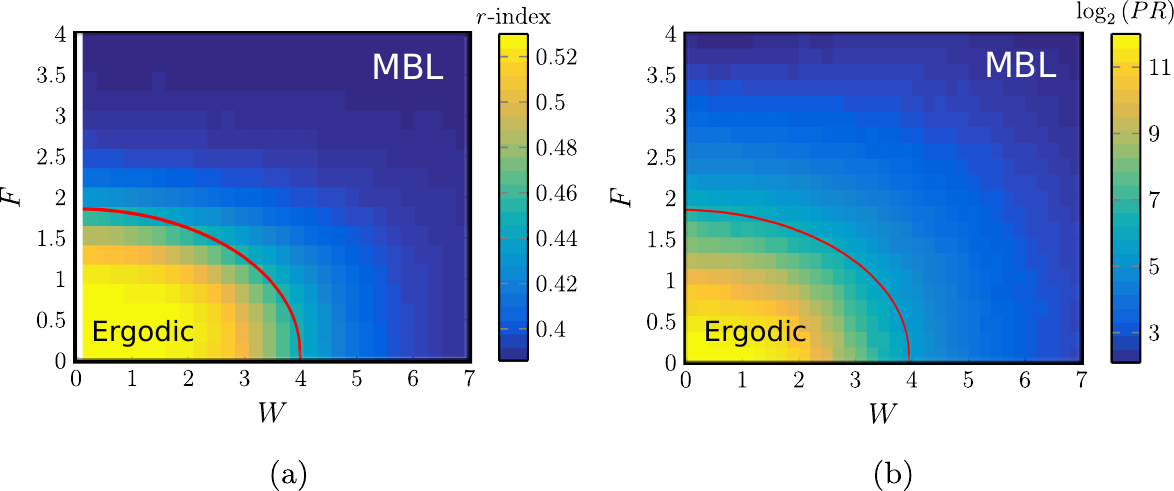}
\caption{These plots constitute the main results of the paper and demonstrate the existence of a potential-gradient induced MBL phase. (a) The $r$-index as a function of disorder and field strength as calculated for the Hamiltonian in \refEq{spin_H} with $L=16$ and $J_0=J_z=1$ (averaged over $125$ realizations). Evidently, a phase boundary exists between a region with $r=0.53$ (Wigner-Dyson) for small values of $W$ and $F$ (the ergodic dome) to a region with $r=0.386$ (Poisson). (b) The averaged participation ratio ($PR=1/IPR$) as a function of disorder and field strength for the same system as in (a). Consistently with the level statistics, inside the ergodic dome the $PR$ is proportional to the Hilbert space dimension ($\mathcal{D}$), while outside the dome it becomes small and independent of $\mathcal{D}$. Notice that in (b) the line $W=0$ is included in the data. In both cases the red line serves only as a guide to eye and is a contour or $r\approx 0.46$. \label{2d_PD}}
\end{figure*}
In this work we show that single-particle-localization that is not necessarily due to disorder, results in a state that is indistinguishable from the MBL state based on the typical tools of assessment.
We analyze the spectral and the dynamical properties of one-dimensional interacting fermions and spins in the presence of both disorder and a linear potential. We show that by considering these two different localizing mechanisms, i.e., disorder ($W$) and linear fields ($F$), one may construct a two-dimensional phase diagram in the $(F,W)$-space which hosts a connected non-ergodic (MBL) phase. We find that above a critical value $F_c$, the MBL phase extends down to the clean limit, i.e., the $W=0$ line.

It is worth mentioning that integrable models, such as the $1$D Heisenberg and transverse field Ising models, are known examples of clean models that fail to thermalize. While these models fail to thermalize, they are sensitive to the existence of small integrability-breaking terms such as disorder or longer range interactions and hopping. In this sense the model we suggest is more generic, since the addition of disorder and/or weak longer range hopping and interactions does not lead to thermalization.

The existence of generic clean models that fail to thermalize may have important implications both theoretically and experimentally. From the theory side, it can simplify dramatically the numerical effort in analyzing these interacting systems. Moreover, the lack of randomness gives hope that the nature of the MBL transition, the emergent conserved quantities and the generalization to higher dimensions may be approached analytically.
From the experimental side the necessity of strong disorder is a major drawback. In intrinsic systems it is not clear whether such strong disorder generically exists. In controlled systems, such as optical lattices, only quasi-random disorder or correlated disorder, e.g. speckle potentials, may be implemented and a repetition over many realizations is needed due to the small size of the systems \cite{oldis1,oldis2}. In stark contrast, linear field (tilt in optical lattices) may be implemented relatively easily and it provides the ability to experimentally realize these systems in a highly reproducible way, and without the necessity of many repetitions. Unlike integrable models, the inevitable existence of unwanted terms such as weak disorder, should not have a dramatic effect on the dynamics.

\section{Background and Model Definition}
\subsection{Bloch localization}
Our ultimate goal is to understand the fate of Bloch localization in the presence of interactions. In this section we briefly review the properties of non-interacting particles in the presence of a uniform force (linear potential).
Consider a $1$D lattice model in the presence of a linear potential,
\begin{equation}\label{BL_H}
H_0=\sum\limits_j t(c^{\dagger}_{j}c_{j+1}+h.c)-Fjc^{\dagger}_{j}c_{j},
\end{equation}
where $c_j$ annihilates a particle from lattice site $j$, $t$ is the nearest neighbor hopping amplitude, and $F$ is the uniform force.
The Hamiltonian can be diagonalized by the following transformation,
\begin{equation}\label{WS_eigs2}
b_m=\sum\limits_j \mathcal{J}_{j-m}\left(x\right)c_j,
\end{equation}
with $\mathcal{J}_n$ being the Bessel functions of the first kind and $x=2t/F$. Under this transformation \refEq{BL_H} becomes,
\begin{equation}\label{BL_HD}
H_0=-\sum\limits_m Fm\,b^{\dagger}_{m}b_{m}.
\end{equation}

Since $|\mathcal{J}_{n}\left(x\right)|<e^{-|n|}$ for $x\ll n$, all the eigenstates are localized for any $F\neq0$.
Each eigenstate, $b^{\dagger}_{m}|vac\rangle$, is localized around site $m$ with an inverse localization length given by $\xi^{-1}\approx 2\sinh^{-1}(1/x)$.

Unlike for Anderson localization, where the localization length is energy dependent (smaller near the middle of the energy band), for Bloch localization case the localization length is an energy independent quantity. Another prominent difference between the two is the form of the density of states, where in the case of Bloch localization the spectrum forms an ordered ladder even deep in the localized phase.

\subsection{Model Definition}
The basic model we wish to analyze concerns the interplay between the two mechanisms of single particle localization (disorder and linear field) and interactions. For that, we consider a $1$D lattice of interacting spinless fermions in the presence of disorder and a uniform force,
\begin{equation}\label{fer_H}
H=\sum\limits_j t(c^{\dagger}_{j}c_{j+1}+h.c)-Fjn_{j}+h_{j}n_{j}+Un_{j}n_{j+1},
\end{equation}
where $c_j$ annihilates a particle from lattice site $j$, $n_j=c^{\dagger}_{j}c_{j}$ is the density, $t$ is the nearest-neighbor (nn) hopping amplitude, $F$ is the uniform force, $h_j\in [-W,W]$ is a random on-site potential with strength $W$ and $U$ is the nn interaction strength.

The above fermionic Hamiltonian may be mapped, via a Jordan-Wigner transformation, into an equivalent spin-$1/2$ chain (Heisenberg),
\begin{equation}\label{spin_H}
H=\sum\limits_j J_0(S^{x}_{j}S^{x}_{j+1}+S^{y}_{j}S^{y}_{j+1})+J_zS^{z}_{j}S^{z}_{j+1}+FjS^{z}_{j}+h_{j}S^{z}_{j},
\end{equation}
with $J_0=2t$ and $J_z=U$ while $F$ and $h_j\in [-W,W]$ defined as before. In the rest of this paper we will analyze the localization and dynamical properties of these Hamiltonians as a function of the interaction strength, force and disorder strength.
Since the particle-number (fermionic model) or the total $S_z$ (spin model) are conserved, we focus our analysis on the half-filled ($S_z=0$) sector. Regardless, the results do not depend much on the specific sector.

\section{Results and Discussion}
\subsection{Level statistics}
A well established signature for the transition from ergodic to non-ergodic dynamics is the level statistics of the many body spectrum. In particular, generic ergodic Hamiltonians belong to the Gaussian Orthogonal Ensemble (GOE) \cite{GOE2,GOE3} and their level-spacings, $\delta_n=\epsilon_{n+1}-\epsilon_n$, typically obey the Wigner-Dyson distribution. On the other hand, for non-ergodic systems the level-spacings typically obey the Poisson distribution. It should be pointed out that in both cases, symmetries may add high level of degeneracies which lead to deviation from the Wigner-Dyson distribution (non-ergodic) and from perfect Poisson distribution (ergodic). Yet, level spacing obtained from symmetry
sectors should not have these additional degeneracies. As in the case of the disordered Heisenberg chain within the sector of zero total magnetization, the transition from ergodic to non-ergodic is accompanied by a transition from Wigner-Dyson to Poisson level statistics \cite{PalHuse}.

Both distributions are often characterized by a single parameter, $r=\left\langle\min(\delta_{n},\delta_{n+1})/\max(\delta_{n},\delta_{n+1})\right\rangle$, which conveniently avoids the need for unfolding the spectrum. For the Wigner-Dyson distribution $r\approx 0.530$ and $r = \ln 4 - 1 \approx 0.386$ for the Poisson distribution.

We diagonalize the Hamiltonian in \refEq{spin_H} for $L=12,14,16,18$ spins using exact diagonalization, with $J_0=J_z=1$ and for different values of $F$ and $W$. In the appendices we show that by employing machine learning techniques, statistics for the $r$-value may be generated from $h_j$ directly without the need of diagonalizing the Hamiltonian.

\begin{figure}[t!]
\centering
\includegraphics[width =\linewidth]{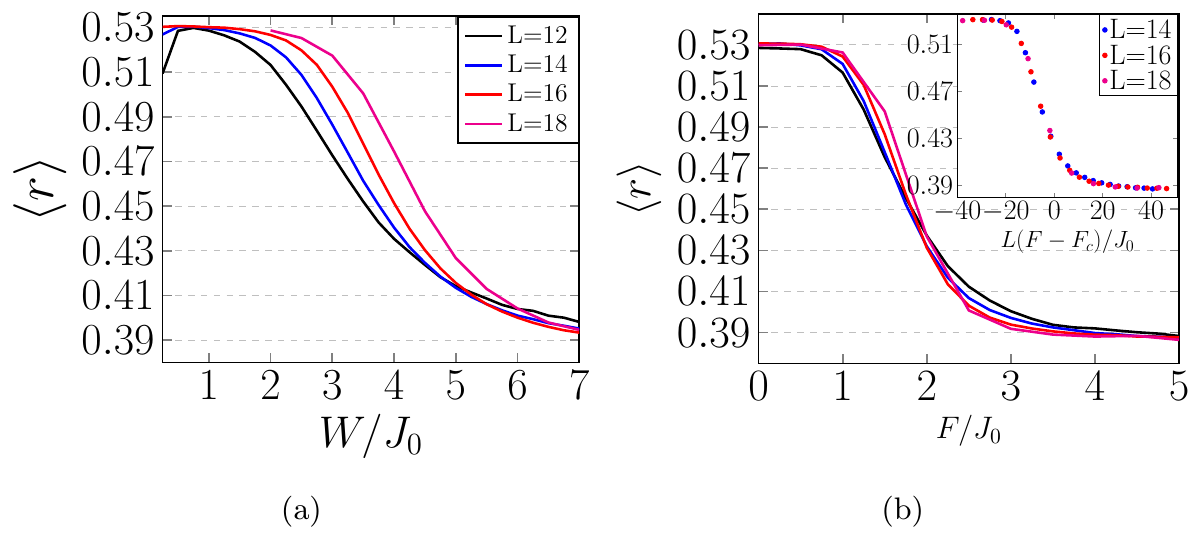}
\caption{The $r$-index as calculated for the Hamiltonian in \refEq{spin_H} with $J_0=J_z=1$ for different system sizes, $L=12,14,16,18$. In (a) the $r$-index is plotted as a function of $W$ for zero linear field. In (b) the $r$-index is plotted as a function of $F$ for a fixed disorder strength $W=0.5$, where in the inset we plotted the data as a function of $L(F-F_c)$ with $F_c= 2.2$.   \label{F_size_S}}
\end{figure}

In \refFig{2d_PD}(a) we show the $r$ value (averaged over different disorder realizations) in the space of $(F,W)$.
We find that the ergodic phase lives in a dome-shaped region near the origin of the $(F,W)$ space.
The line $F=0$ corresponds to the often discussed MBL transition near the critical disorder strength $W_c$. As $F$ increases, the value of $W_c$ decreases. Above a critical value of $F$, the critical disorder appears to go to zero and the non-ergodic phase appears also in the clean non-disordered limit.

In \refFig{F_size_S} we show the $r$ value for different system sizes as a function disorder (zero field) and as a function of the field (for a fixed weak disorder). The critical values may be extracted by finite size scaling through a scaling collapse. The case of zero field was analyzed in several works \cite{HuseOgan,finite_size_sc1,finite_size_sc2,finite_size_sc3} in which the critical disorder was found to be in the range $W_c\sim 7.5\pm 0.5$ (notice a factor of $2$
due to a different definition of the spin matrices). For the weak disorder case we plot the data, \refFig{F_size_S} (inset), as a function of $L^{1/\nu}(F-F_c)$. We find that the critical exponent is $\nu\approx 1$ and the critical field is $F_c\approx 2.2$, for which the data collapse on one curve. In the appendices we provide more details regarding the finite size scaling, and show that the above results are not sensitive to integrability-breaking terms such as next-next-nearest-neighbor hopping and interactions.

Notice that in this part, we always considered $W>0.2$, since for small enough disorder, small systems behave as clean systems which leads to symmetry related degeneracies in the spectrum.

\subsection{Inverse participation ratio}
Analyzing level statistics of clean systems requires a separation of the Hilbert space into momentum sectors, since degeneracies due to symmetries have to be removed. For finite systems and below a critical disorder strength, the system behaves similar to a clean system. Therefore, the level statistics becomes a less reliable measure for small disorder strengths since degeneracies start to appear due to the emergence of translation symmetry. A quantity which is less sensitive to symmetries is the inverse participation ratio (IPR). The IPR is also a measure of the long-time return probability of arbitrary initial states. To see that, consider the return probability of a state $|\psi_0\rangle$,
\begin{equation}\label{return}
P(t)=\left|\langle\psi_0|\hat{U}(t)|\psi_0\rangle \right|^2,
\end{equation}
where $\hat{U}(t)$ is the time evolution operator. The state $|\psi_0\rangle$ may be expanded in terms of the Hamiltonian eigenstates, $|\psi_0\rangle=\sum_n c_n|\phi_n\rangle$ which allows to write
the IPR (long-time limit of the return probability) as,
\begin{equation}\label{IPR}
IPR=\lim\limits_{T\to\infty}\frac{1}{T}\int\limits_{0}^{T}dtP(t)=\sum\limits_{n,m}|c_n|^2|c_m|^2\delta_{\epsilon_n,\epsilon_m}.
\end{equation}
In the absence of degeneracies, \refEq{IPR} becomes $IPR=\sum_{n}|c_n|^4$. Clearly, if the initial state is an eigenstate then $IPR=1$, while if the initial states is an equal-superposition of all the eigenstates then $IPR=1/\mathcal{D}$, where $\mathcal{D}$ is the Hilbert space dimension which generically is exponential in the system size. In the following we average the $IPR$ over different initial states which we choose to be eigenstates of some local operators, e.g., $s_{j}^{z}$. For ergodic systems, the $IPR$ should be exponentially small in the system size and the system should lose its memory of the initial state. In stark contrast, in the localized phase the $IPR$ converges to a positive system size independent constant.

In \refFig{2d_PD}(a) we present the averaged and normalized participation ratio, $\langle PR\rangle=\mathcal{D}/IPR$, in the space of $(F,W)$. While the IPR is a smooth function, there is a transition between a region where the IPR is exponentially small to a region where the IPR is independent of system size. These regions agree with the results obtained in the previous section. Here also the line $W=0$ behaves in a similar way (c.f. \refFig{2d_PD}(b)), where the IPR becomes independent of system size as a function of $F$.

\subsection{Dynamics and experimental measurables}
\label{subsec:Dynamics}
The distinction between ergodic and non-ergodic dynamics is well-captured by the level-statistics and the participation ratio. Yet both these measures are hard to access in experiments.
As shown in \citeRefs{exp2,exp3,exp5}, the nature of the dynamics is examined by tracking the dynamics of an initially prepared out-of-equilibrium density configuration. We numerically show that the existence of a linear field prevents thermalization. For concreteness, we consider a similar out-of-equilibrium initial state as in \citeRef{exp2}.
\begin{figure}[t!]
\centering
\includegraphics[width =\linewidth]{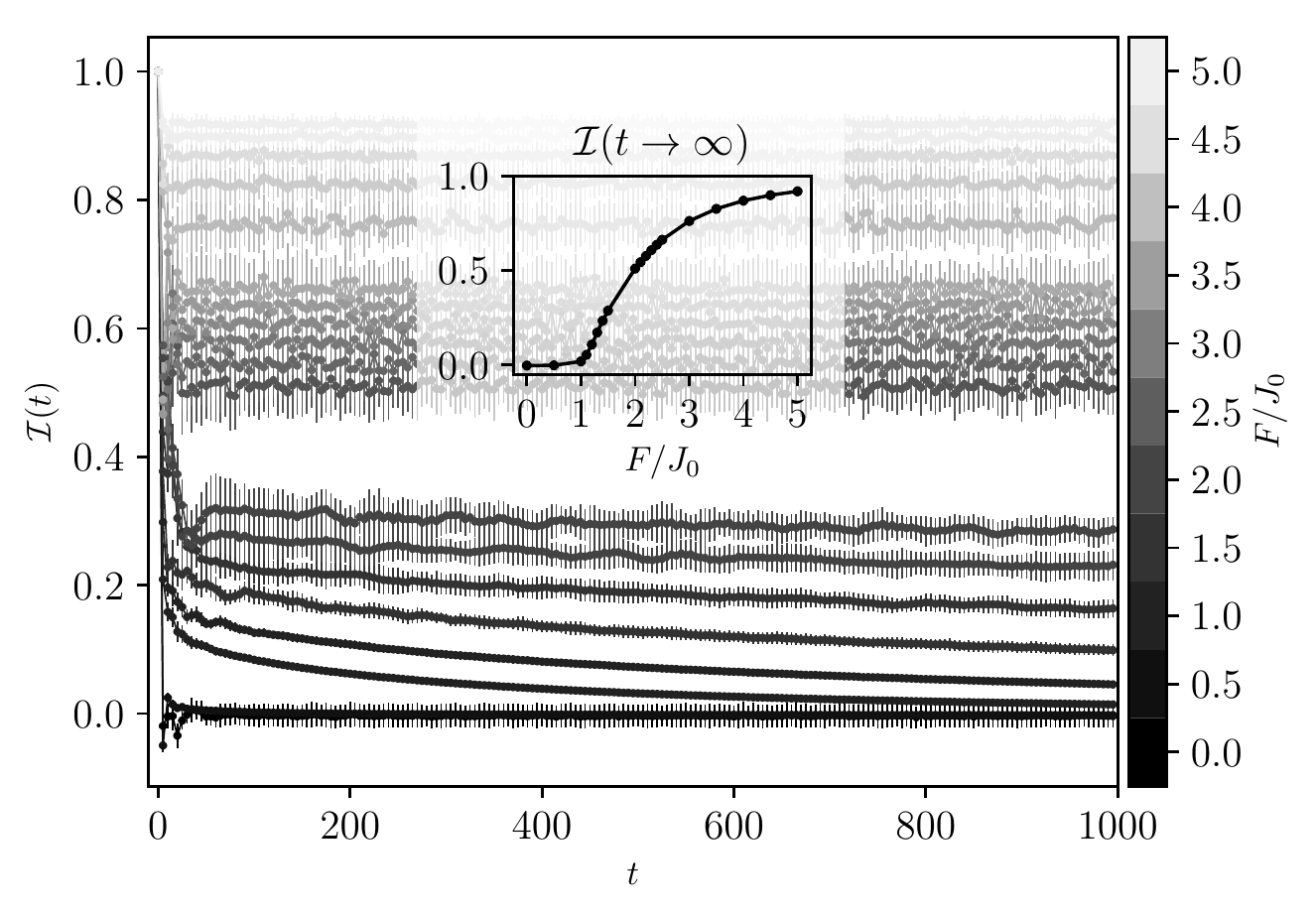}
\caption{The imbalance $\mathcal{I}$ as a function of time for different field strength and for fixed weak disorder $W=0.2$, where $L=24,\,J_z=J_0=1$. At $t=0$, the imbalance for each field strength starts at $\mathcal{I}(t=0) = 1$. For field strengths above (and including) $F=2.0$ we can not identify decaying behavior at these time-scales. Error bars show statistical variance over $32$ realizations of disorder. The inset shows the long time limit of the imbalance as a function of the field (averaged over the last 50 timesteps). Below a critical value $F\lesssim F_c$ the long time limit of $\mathcal{I}$ tends to zero, while above that value the long time limit tends to a finite value that increases with the field. Notice that some of the lower $F$ curves have not yet reached their final value. \label{Imb}}
\end{figure}
The system is prepared in a anti-ferromagnetic configuration (or charge density wave for the fermions), where the spins on odd sites point down (empty) and on even sites point up (full). We then track the time evolution of the odd-even imbalance, $I=(S_{z,\uparrow}^e-S_{z,\uparrow}^o)/(S_{z,\uparrow}^o+S_{z,\uparrow}^e)$.
We use a numerical method based on Krylov-subspaces via a re-orthogonalized Lanczos implementation to do so (see the appendices for more information).

In the inset of \refFig{Imb} we show the resulting long time limit as a function of $F$.
Below a critical value $F\lesssim F_c$ the long time limit of $\mathcal{I}$ tends to zero, while above that value the long time limit tends to a finite value that
increases with the field. Computational costs limit the available times we can access, and we remark that not each of these curves have converged yet. Extrapolating
the available curves will move the estimated critical field to higher values than suggested by the inset.

In ergodic systems, $\mathcal{I}$ is expected to decay to zero with a typical relaxation time $\tau$.
We show that while indeed this is the case when the linear field is small, both for the clean case and for weak disorder, beyond a critical field strength,
the long time limit of $\mathcal{I}$ is different from zero. In \refFig{Imb} we show the imbalance $\mathcal{I}$ in a system of $24$ spins (sites) as a function of
time for different values of the field $F$ and for a fixed weak disorder strength ($W=0.2$).
It is worth noticing that energetics gives an upper bound to this relaxation process. In the fermionic language, the charge density wave (CDW) configuration and the uniform configuration differ in their dipole moment $D = \sum_j jn_j$ by an extensive amount $\Delta D=N/4$. In the absence of a field, the many-body bandwidth of the Heisenberg model with all $J=1$ is $\log(2) N$. Hence, in the presence of a field, if $F\Delta D>\log(2) N$ or $F>4\log(2)\approx 2.77$
the CDW configuration can not evolve into a uniform configuration at any time. In practice, the critical field obtained from the the level statistics is around $F\sim 2.2$ while it seems that the dynamics suggests a slightly lower value (notice that not all of the imbalance curves have converged yet).
Yet, while the the true critical field (if it exists) should limit the dynamics of all processes, specific processes like the one considered here may show non-ergodic
dynamics at lower values. Moreover, one may consider the presence of a pre-localized phase (or pre-thermal for that matter) that appears in the dynamics.
Our numerical data cannot confirm or disprove the existence of such a phase. An analysis of larger systems, and more importantly, much longer times, may
resolve that issue.

\section{MBL in two-dimensions}
The lesson we learned about the effect of interaction on the Anderson localized (AL) phase in $1$D can not be trivially extended to higher dimensions. The nature, and even the existence, of a many-body-localized phase in $D>1$ is a hotly debated subject.
While theoretical works \cite{mbl2d1,mbl2d2,mbl2d3} showed that locally thermal regions in systems with true random disorder can destabilize the MBL phase in two dimensions, experimental works \cite{exp2,exp4,exp5} have shown indications for such a phase in $D>1$.

Similar questions may be posed in the context of the uniform field as a cause for single particle localization. In stark contrast to the AL phase, this phase is not sensitive to rare regions. In particular, if the field is applied at an irrational angle, the field is non-zero along
all lattice directions. This field can indeed be arbitrarily small for specific lattice directions, but can be chosen such that lattice sites along directions at which the field is below the critical
value are separated by multiple hops. Since each of these hopping processes has a component against a strong field, and since the bare interaction is local, both the
effective hopping coefficient and the effective interaction along these directions may be extremely small. How these renormalized hopping coefficients and interactions scale with the field along these direction, and
whether it is possible to choose the field such that along
each lattice direction the field is larger than the $1$D critical field, is an interesting question worth further investigation.
Additionally, along these same lattice directions the linear potential is not perfect and can be regarded as a combination of a linear field and weak quasi-periodic disorder.
This quasi-periodic disorder may also help the localization along directions where the field is small.
Finally, the absence of rare regions (which are a main reason to exclude MBL in high dimensions~\cite{mbl2d2}) may help the survival of the non-ergodic
behavior in the thermodynamic limit.

\begin{figure}[t!]
\centering
\includegraphics[width =\linewidth]{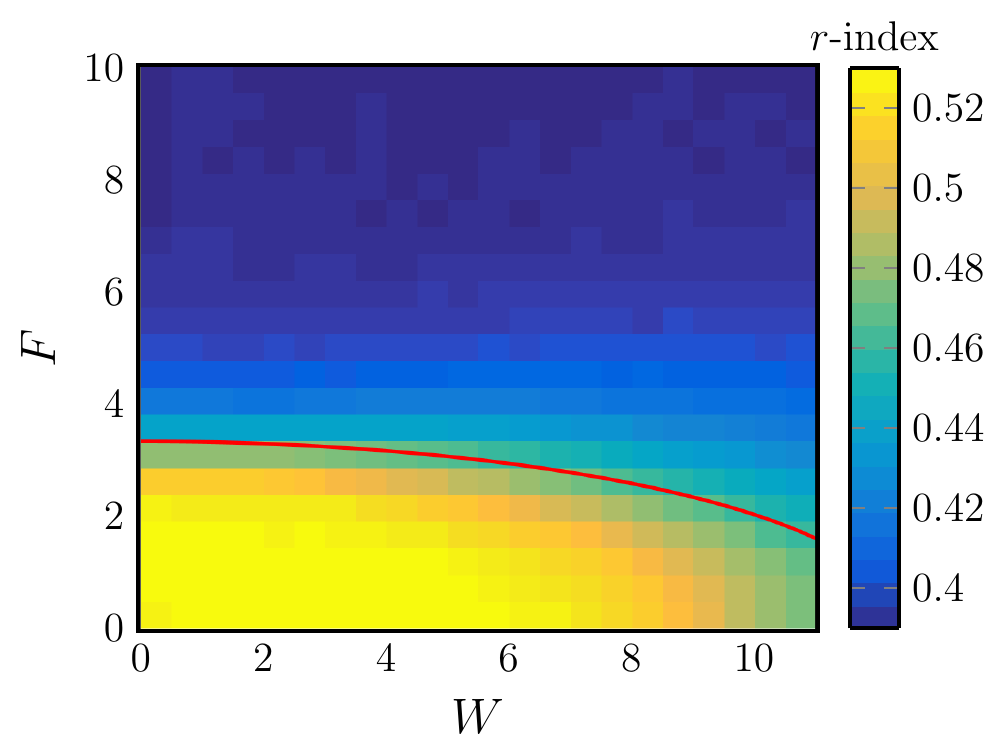}
\caption{The level statistics ($r$-value) as a function of the field strength, $F$, and disorder strength, $W$, for a disordered $2$D system of $4\times 4$ spins with an incommensurate force $\textbf{F}=F(\sqrt{2},1)$ (averaged over $32$ realizations). The red line is a guide to the eye and is given by a contour of $r\approx 0.46$.\label{2D_ls}}
\end{figure}

To further speculate on the existence of MBL in 2D, \refFig{2D_ls} shows the level statistics (r index) of a $2$D Heisenberg model as a function of the uniform force $\textbf{F}=F(\sqrt{2},1)$ and disorder.
Similar to the $1$D case, we see a clear transition from a Wigner-Dyson distribution to Poisson distribution. Since we are restricted to very small system sizes ($4\times4$ lattice), these results should not be taken as a claim
of the existence of a two dimensional MBL phase. However, we hope that these ideas will stimulate further works in this directions.

\section{Dipole moment analysis}
Single-particle Wannier-Stark localization may be thought of in terms of energetic constraints imposed by the field.
In the many-body case, one may wonder whether interactions can help overcome energetic constraints by reordering of particles.
Heuristically, a many-body configuration can be described by its dipole moment $D$ (see section~\ref{subsec:Dynamics}), which in the presence of a field $F$ is
associated with an energy $FD$. In order for such a configuration to evolve into a different configuration with dipole moment $\tilde{D}$, the internal structure of
the system, i.e., hopping and interaction, must be able to supply the energy difference $F(D-\tilde{D})$. This condition is captured in the dipole moment
structure of the eigenstates. Clearly, in the infinite field limit of our model, the dipole moment becomes an exactly conserved quantity.
In that case an analogy with \citeRef{Nandkishore} can be made, where it was shown that non-ergodic dynamics arises in a one dimensional
random quantum circuit model which is constrained to conserve both a $U(1)$ charge and the dipole moment of this charge.

The main question is to what extent the dipole moment may be considered as a conserved quantity for finite fields.
In the appendices we show the results of exact diagonalization of a half-filled fermionic system where each point represents an
eigenstate in the space of energy and dipole-moment. As expected, in a given energy window and for large field the many body wave functions have well-defined dipole moment. Each dipole moment sector is further split into subsectors of doublon (occupied neighboring sites) number.
Hence the dynamics is effectively restricted to preserve the initial dipole moment and the initial doublon number, which is predicted to yield non-ergodic dynamics~\cite{Nandkishore}.
For a weak field however, this is not the case. The eigenstates in a given energy window span a range of dipole moments and doublon numbers.
Around the critical field, we observe that while the eigenstates in a given energy have a finite spread in the dipole moment, the different sectors become distinct
and the integer part of the dipole moment behaves as a conserved quantity.
Beyond this critical field we also observe a separation into the subsectors of doublon number. While it is hard to pinpoint the exact value of the transition using this approach,
the transition can be bounded and is consistent with the value we obtained from the level spacing statistics.
\section{Conclusions}

In this work we analyzed the effect of interactions on single particle localization that arise both from disorder, $W$, and from the existence of linear potentials $F$. With that, we showed that the notion of a many-body localized (MBL) phase may be generalized also to a class of clean (non-integrable) systems. In particular, we find that a phase boundary in the space $(F,W)$ exists, beyond which the resulting phase fails to thermalize. We find that, unlike in clean integrable models, this non-ergodic phase is stable to perturbations, and shares all the familiar fingerprints of the well studied MBL phase in the presence of disorder.

The existence of such a phase demonstrates that randomness is not an essential ingredient for the emergence of stable non-ergodic interacting phases.
Such a conclusion may have an impact on the realization of these non-ergodic phases.
Unlike disorder potentials, linear potentials are relatively easy to implement, and are highly tunable and may be controlled dynamically.
The ability to realize stable and generic non-ergodic phases is an important step toward the realization of quantum memory devices that may store information for long times.
Moreover, the lack of randomness and the low sensitivity to dimensionality may render these systems more accessible to a further theoretical investigation, both numerically and analytically.
It came to our knowledge that simultaneously to our work, the entanglement-entropy grows in the presence of linear field has been studied in \citeRefs{mblnodis4}. The results presented in \citeRefs{mblnodis4} are in agreement with the conclusions we presented in this work.

\begin{acknowledgments}
E.v.N. gratefully acknowledges financial support from the Swiss National Science Foundation through grant P2EZP2-172185. G.R. is grateful to the ARO MURI W911NF-16-1-0361 ``Quantum Materials by Design
with Electromagnetic Excitation'' sponsored by the U.S. Army as well as the Packard Foundation. We are grateful for support from the IQIM, an NSF physics frontier center funded in part by the Moore Foundation.
\end{acknowledgments}

\def\CC{{C\nolinebreak[4]\hspace{-.05em}\raisebox{.4ex}{\tiny\bf ++}}}
The \CC $\,$ code we developed for this is available online at \url{https://www.github.com/everthemore/krylov-cpp}, and the data for Fig.~\ref{Imb} is available at \url{https://data.caltech.edu/records/1089}.

\appendix
\section{Data augmentation using machine learning}
The different disorder realizations we study in this manuscript differ only in the values for the on-site potentials. Given the on-site potentials, there exists a procedure that results in the value for the $r$-statistics. Namely, one builds the corresponding Hamiltonian matrix and diagonalizes it to obtain the eigenvalues $\epsilon_n$. The r-statistics is obtained by looking at neighboring eigenvalue differences $\delta_n=\epsilon_{n+1}-\epsilon_n$ and computing the ratio $r=\left\langle\min(\delta_{n},\delta_{n+1})/\max(\delta_{n},\delta_{n+1})\right\rangle$ as discussed in the main text.
\begin{figure}[t!]
\centering
\includegraphics[width =\linewidth]{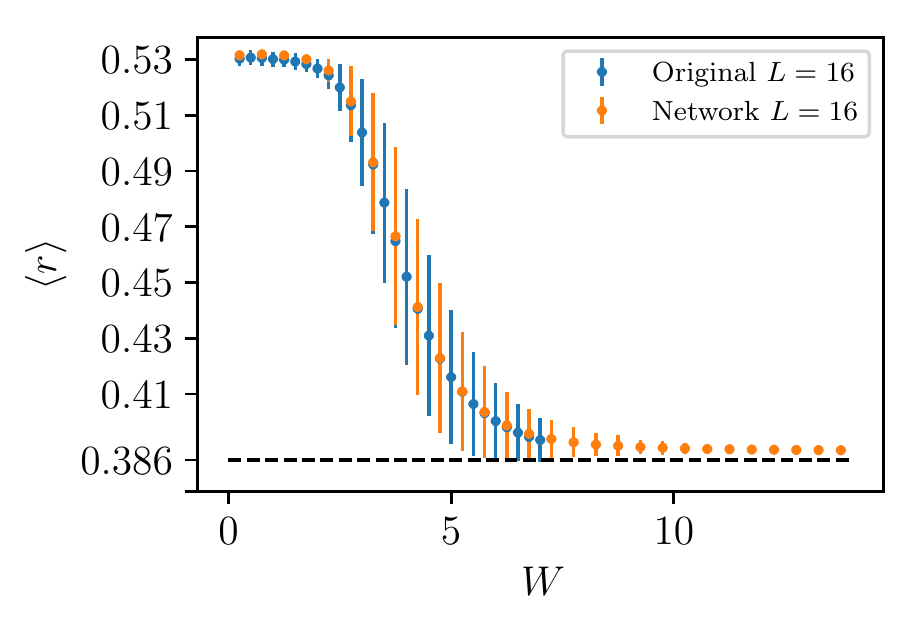}
\caption{The original $L=16$ data and the machine learned map from the disorder realization $h_1$ through $h_{16}$ to the resulting $r$-statistics. With the network we are able to generate considerably more realizations ($10^6$ versus $10^3$) in a much shorter timespan, provided that the network is capable of learning and generalizing. The sigmoid output neuron rather than linear for optimizing the mean-squared-error ensures convergence of the output as a function of $W$. Error bars indicate the standard deviation over the number of realizations, and the black dashed horizontal line indicates the Poissonian $r$-value of $\ln 4 - 1$. \label{fig:MLMBL}}
\end{figure}

Here, however, we ask whether or not a more direct (approximate) map exists from the on-site potentials to $r$. Rather than trying to explicitly construct it, we attempt to train a neural network to perform this map for us. Hence we generate a large data-set of pairs $(\mathbf{h},r)$, where $\mathbf{h}$ is a vector of the on-site potentials augmented with the value of $W$ from which they were drawn, and $r$ is the resulting $r$-statistics for this particular realization. These serve as the input and output respectively for the machine learning model.

Provided that such a mapping exists and that the network is capable of learning it, the resulting network can be used to generate more $r$-values by using it to predict on more realizations. This allows one to generate statistics much faster compared to running the full exact diagonalization.
It must be noted that this procedure cannot take away the inherent statistical uncertainty due to the finite size of the system. Particularly, for disorder strengths near the transitions point, the exact $r$-values of systems with different realizations drawn from the same distribution, lie within a relatively large window. As the system becomes larger  this window becomes smaller. For example, already by including a few hundreds of realizations, for $L=16$, the error bars near the transition are dominated by the intrinsic finite size effect and cannot be improved by adding more realizations.

In Fig.~\ref{fig:MLMBL} we demonstrate the above procedure for the $L=16$ data, for which the data-set consists of $\sim 15$k entries
($25$ values of $W$ spread over $\sim 550$ realizations). We split off $10\%$ of the data as a validation set, and train a network with the
following architecture. First, two convolutional layers with $32$ filters and kernel sizes $6$ and $3$, followed by a maximum pooling of size 3.
Then a convolutional layer with 64 filters and kernel size 2, followed by a global average pooling. Next, two
fully connected sigmoid layers with 256 and 128 neurons respectively, and dropout 0.5. And finally an output layer with a single sigmoid neuron.

\begin{figure}[t!]
\centering
\includegraphics[width =\linewidth]{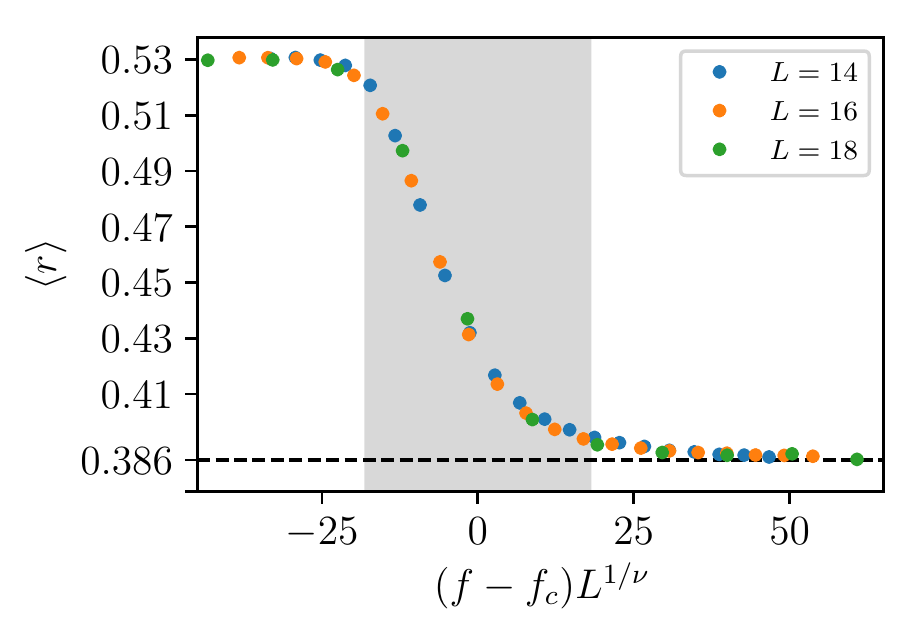}
\caption{Collapse of the $W = 0.5$ data for system sizes $L=14,16,18$, as a function of the field strength $f$. The collapse is obtained by rescaling the fields according to $f \to (f-f_c)L^{1/\nu}$ with $f_c = 2.08$ and $\nu = 0.952$. The gray area indicates the width $w$ that was used to make the curves collapse, and is the width at which the collapse is most stable against inclusion or removal of the $L=12$ data.\label{fig:FSSfield}}
\end{figure}

We train the network with the Adam~\cite{kingma} optimizer to minimize the mean-squared-error loss function, and achieve a validation loss of $\sim 2 \cdot 10^{-5}$ in 100 epochs of batchsize $32$. In our experiments, we have found no particular reason for the above network to work better than others, but we found that considerably simpler networks (e.g. just fully connected layers) converge much slower. For the purpose of extracting the mapping, our chosen network might be hard to interpret. It would be an interesting research direction however to see if the approximate mapping can be extracted from a network, or whether a single network can be trained on different system sizes to extract finite size behavior. Both would potentially allow predictions to be made on larger system sizes than trained on, although further investigation into this question is required.

\section*{Finite size scaling}
In this appendix we discuss the transition from the ergodic to the non-ergodic phase as a function of the linear field $f$. To do so, we fix $W = 0.5$ and perform a finite size scaling analysis attempting to collapse the curves for different system sizes. We consider a universal function $g\left( \left(f-f_c\right)L^{1/\nu} \right)$ for the $r$-statistics, and optimize the parameters $f_c$ and $\nu$ so that the rescaled $r$-statistics curves for the different sizes collapse.

Each of the curves is first rescaled with proposed $f_c$ and $\nu$ after which we use spline interpolation to numerically minimize the cost function $C(f_c,\nu) = \sum_{i < j}\int_x (y_i(x) - y_j(x))^2$, where $i,j$ both run over system sizes $L = 12,14,16,18$ and $y_i(x)$ represents the spline-interpolated data. The integration regime $x$ is taken to be centered around the transition (i.e.~$x = 0$) and has a width $2w$ that we vary to obtain statistics on $f_c$ and $\nu$. In the collapse including the system size $L = 12$ data, the $L = 12$ curve is consistently the most off. In the spirit of Ref.~\cite{finite_size_sc1} we consider the width $w$ for which the extracted parameters are least sensitive to the inclusion/removal of the $L=12$ data. This results in the parameters $f_c = 2.08 \pm 0.10$ and $\nu = 0.952(5)$.
The resulting collapse for this set of parameters is shown in Fig.~\ref{fig:FSSfield}.

\section{Choice of gauge for numerics}
\begin{figure}[t!]
\centering
\includegraphics[width =\linewidth]{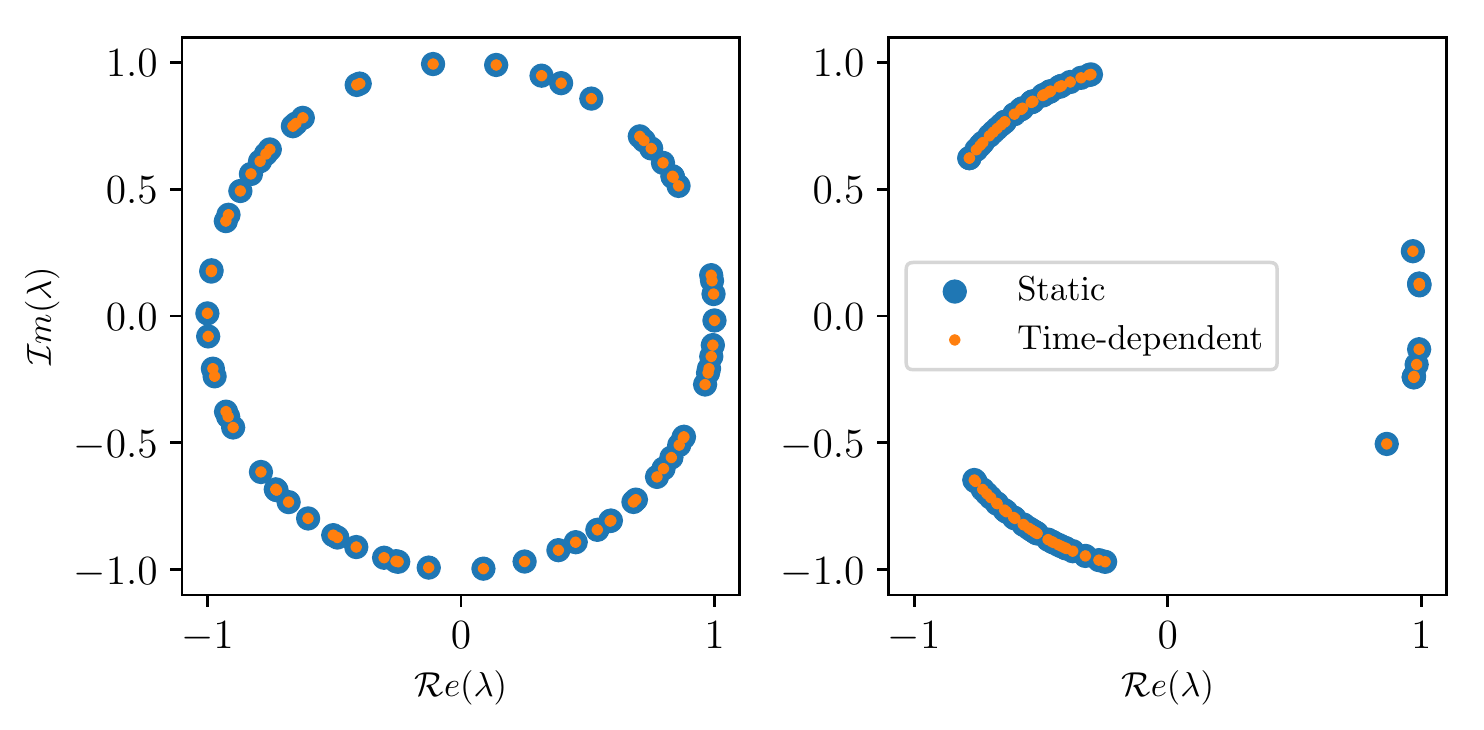}
\caption{The (real and imaginary parts of the) spectrum of $U(T)$ computed in the static gauge (larger blue dots) versus the spectrum of $U(T)$ computed in the time-dependent gauge (smaller orange dots),
for field strengths $F=0.5$ (left panel) and $F=3.0$ (right panel). The difference between the spectra $\sum_{i}|\lambda_{i,\textrm{static}} - \lambda_{i,\textrm{time}}|$ is of
order $\textrm{d}t$ used to calculate the latter. \label{fig:gaugecomparison}}
\end{figure}

We chose to work with a time-independent Hamiltonian for which the linear field is added via the dipole term, rather than as a time-dependent phase factor for the hopping.
This interpretation brings with it the potential issue of having an infinite energy difference between the endpoints of our system as one scales up to the thermodynamic limit.
The physics in these two gauges is evidently invariant, but since we consider (rather small) finite size systems the infinite energies are not a concern. Working in the
time-independent gauge is numerically considerably more convenient, since the time evolution operator over a period $T$, i.e. $U(T)$, can be constructed by a single exponentiation
through as $U(T) = \exp \left(-i H_{\textrm{static}} T\right)$. For the time-dependent case, one would have to compute the time ordered integral $U(T) = \mathcal{T} \exp \left(-i\int \textrm{d}t H(t)\right)$ by
breaking it down into many small $\textrm{d}t$-sized steps and exponentiating $H(t)$ for each. The resulting differences in the spectra $\lambda_{i,\textrm{static}}$ and $\lambda_{i,\textrm{time}}$ are only of the order of
$\mathcal{O}(\textrm{d}t)$. An interesting phenomenon for future investigation is the observed clustering of the eigenvalues for field $F > F_c$, shown in \refFig{fig:gaugecomparison}.

\section{Sensitivity to integrability-breaking terms}
We now consider an extended version of Eq. 5 of the main text,
\begin{align}\label{fer_H_ex}
H=&\sum\limits_j t(c^{\dagger}_{j}c_{j+1}+h.c)-Fjn_{j}+h_{j}n_{j}+Un_{j}n_{j+1}\\ \nonumber
&+\zeta\left(c^{\dagger}_{j}c_{j+2}+h.c+n_{j}n_{j+2}\right).
\end{align}
\begin{figure}[t!]
\centering
\includegraphics[width =\linewidth]{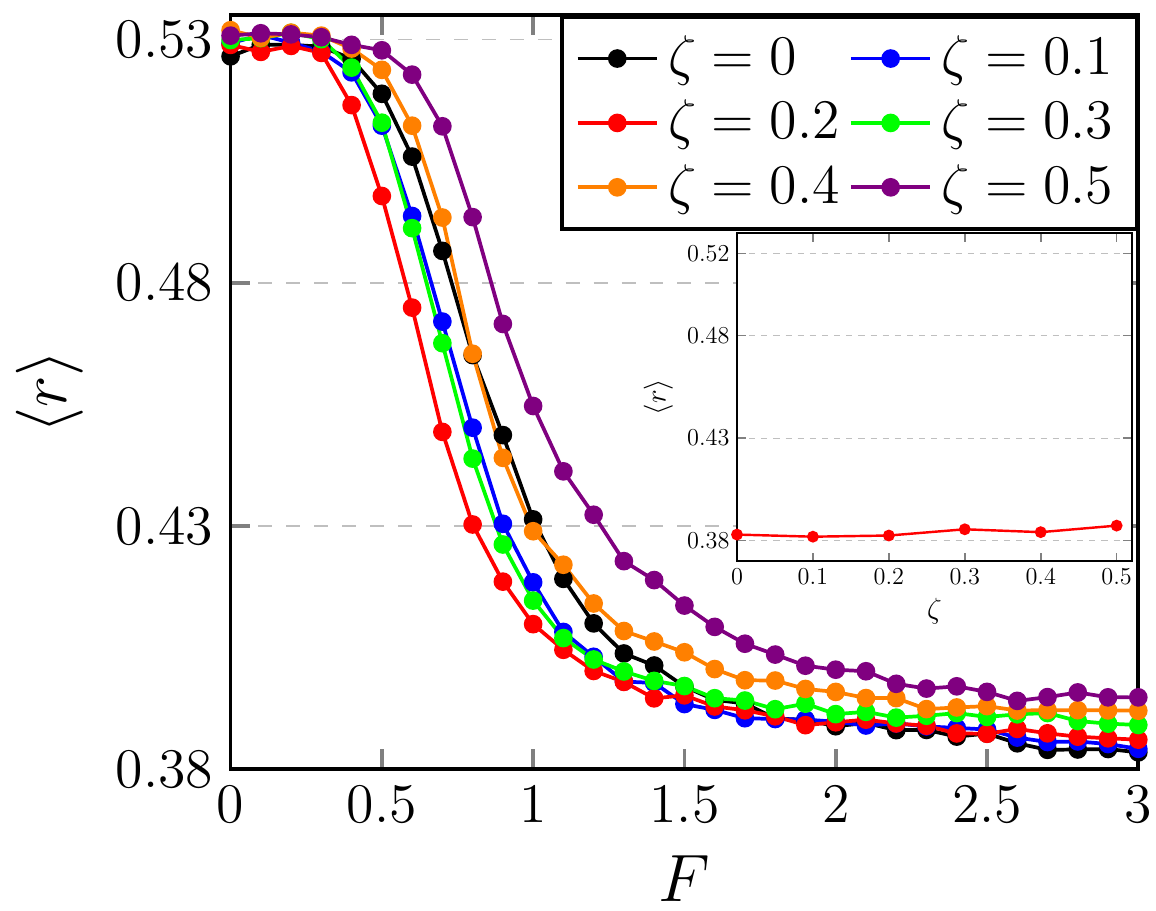}
\caption{The level statistics ($r$-index) as a function of the linear field for different values of the integrability-breaking strength, $\zeta$. The calculation was done for a system of $14$ sites (half-filled) with a fixed weak disorder $W=0.2$ (averaged over $50$ realizations),$t=1/2$ and $U=1$. Inset: the $r$-index of a clean system of $16$ sites with fixed field $F=3$ as a function of $\zeta$. \label{int_break}}
\end{figure}

In the absence of both disorder and linear field, the above model is integrable for $\zeta=0$. We show that also in the presence of the integrability-breaking terms, the application of linear field (with or without disorder) leads to a transition from a Wigner-Dyson level statistics (ergodic) to a Poisson level statistics (non-ergodic). While the value of the critical field depends on $\zeta$ and the disorder strength, the qualitative behavior is indifferent to these terms. In \refFig{int_break} we show the $r$-index as a function of the linear field strength. Different curves represent different values of $\zeta$.

\section{Time-evolution using the re-orthogonalized Lanczos algorithm}
In this appendix we discuss algorithmic details of simulating the time-evolution of a wavefunction using a Krylov-subspace method.
In particular, we have used the so-called Lanzcos algorithm with re-orthogonalization to obtain the results presented in Fig. 3 of the main text.

We wish to numerically perform the time-evolution of a wavefunction, i.e., to compute $|\psi(t_0 + t)\rangle = e^{-i H t} |\psi(t_0)\rangle$.
To do this exactly would require the full diagonalization of the Hamiltonian $H$, which becomes impossible for large system sizes due to memory requirements.
An improvement can be made by using a sparse matrix implementation of the Hamiltonian and iteratively simulating
\begin{equation}
	|\psi(t_0 + \textrm{d}t)\rangle = e^{-i H \textrm{d}t} |\psi(t_0)\rangle
\label{eq:timeevo}
\end{equation}
for small time-steps $\textrm{d}t$.
A naive implementation of this iterative algorithm quickly accumulates numerical errors and becomes unstable, however, a more stable variant can be constructed using Krylov-subspaces~\cite{VanLoan}.
A Krylov-subspace of dimension $m$, $\mathcal{K}_m(H,|\psi\rangle)$, is defined as the span of the vectors $\left( |\psi\rangle, H|\psi\rangle, H^2|\psi\rangle, \ldots, H^{m-1}|\psi\rangle \right)$.
The vector $|\psi(t_0 + \textrm{d}t)$, after expanding the exponent on right-hand side of Eq.~\ref{eq:timeevo}, is approximated well by a vector in this Krylov subspace.

The vectors in $\mathcal{K}_m(H,|\psi(t_0)\rangle)$ first need to be orthonormalized (discussed in more depth shortly), after which we store them as the columns of a new matrix $Q_m$ of dimension $\mathcal{N}\times m$, where $\mathcal{N}$ is the size of the Hilbert space.
After obtaining $Q_m$, we project the Hamiltonian into the spanned subspace to obtain $h_m = Q_m^\dagger H Q_m$. This is a much smaller $m\times m$ matrix that can be easily exponentiate, and allows us to compute
\begin{align}
	|\psi(t_0 + \textrm{d}t)\rangle &= e^{-i H \textrm{d}t} |\psi(t_0)\rangle \nonumber \\
			&\approx \textrm{the first column of } Q_m e^{-i h_m \textrm{d}t}.
\end{align}
In all of the above, the Krylov subspace dimension $m$ can either be systematically increased until convergence is obtained, or changed adaptively during the orthogonalization procedure described next.

\begin{figure}[t!]
\centering
\includegraphics[width =\linewidth]{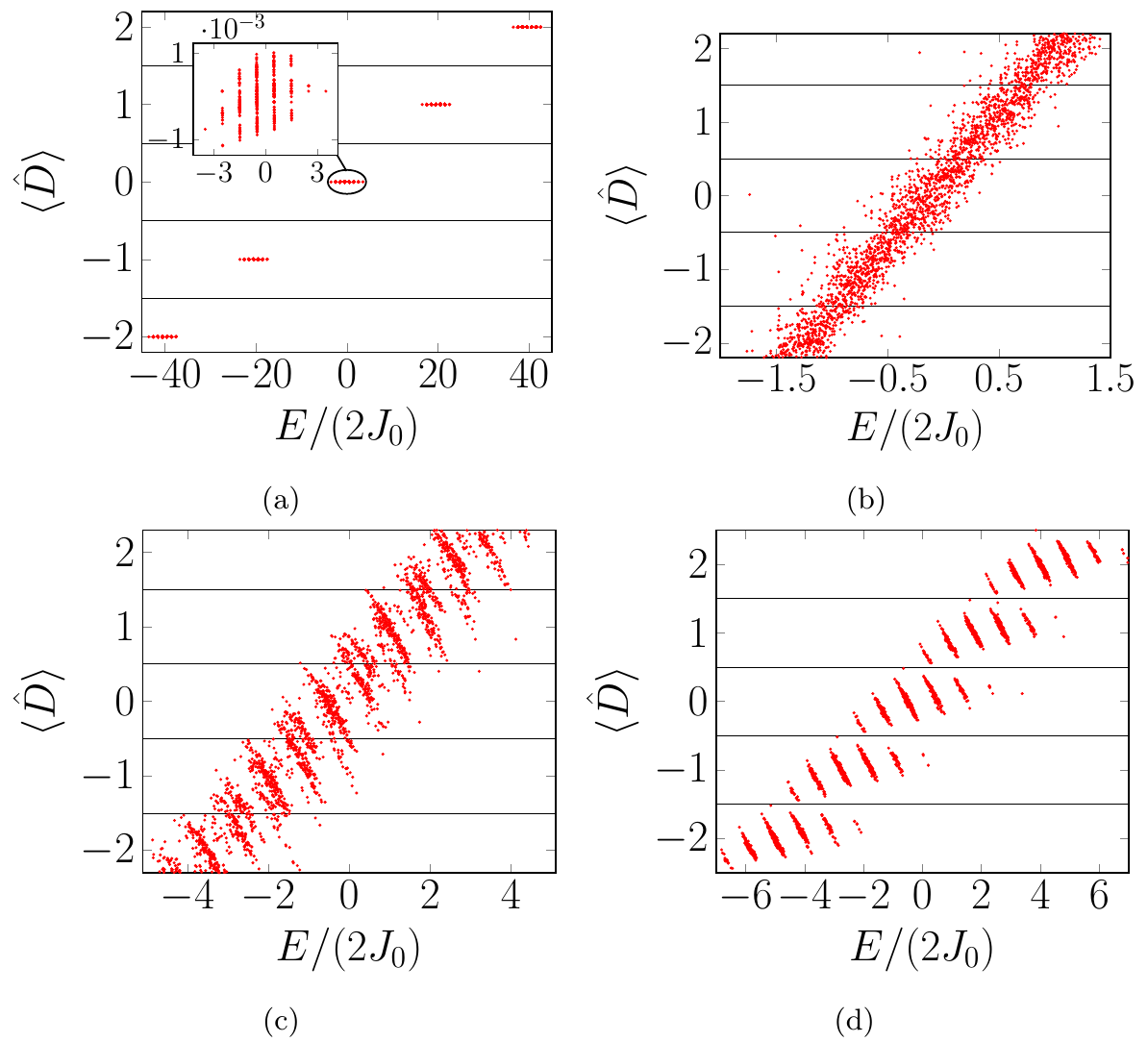}
\caption{Eigenstates dipole moment (expectation value) as a function of their energy for $16$ sites half filled chain with $J_0=1/2$, $U=1$ and different fields: (a) $F=20$, (b) $F=0.5$, (c) $F=1.5$, (d) $F=2.2$. For presentational reasons we show only the expectation value of the dipole moment and omit the fluctuations. The mean fluctuations are (a) $\sigma_F\approx0.04$, (b) $\sigma_F\approx3.2$, (c) $\sigma_F\approx1$, (d) $\sigma_F\approx0.6$. Above a critical field, the eigenstates in a given energy window have a well define dipole moment which restrict the dynamics. \label{dipole}}
\end{figure}

The numerically most challenging step in this algorithm is obtaining the orthonormalized set of vectors for $Q_m$ from $\mathcal{K}_m(H,|\psi(t)\rangle)$.
A standard Gram-Schmidt procedure for orthonormalizing a set of vectors loses the orthogonality between successive vectors simply due to rounding errors (i.e. finite precision of floating point numbers). The modified
Gram-Schmidt procedure does considerably better, but we have found it insufficient for our purpose. The set of vectors we wish to orthonormalize is a special set, in which each vector is generated from the previous
one by application of a matrix. This means we can generate the vectors during the Gram-Schmidt procedure instead of having them given to use beforehand. This small but important difference leads to this algorithm
often being called the Arnoldi method. The resulting projected matrix is in general an upper Hessenberg matrix (upper triangular plus the first lower off-diagonal). If the matrix is
Hermitian as it is in our case, the projected matrix is therefore tri-diagonal. The Arnoldi algorithm with a Hermitian matrix is called the Lanczos algorithm, and provides an improvement in terms of computational effort.

Regardless of using modified Gram-Schmidt, Arnoldi or Lanczos, the orthogonality between successive vectors is gradually lost. A significant improvement, at computational cost of course, can be made by simply re-orthogonalizing
the set of obtained (semi-)orthogonal vectors. It turns out that for the re-orthogonalization ``twice is enough'' for non-singular cases~\cite{Giraud2005}. For the numerics presented in Fig. 3 of the main text, we have checked the
convergence of the curves with respect to the timestep $\textrm{dt}$ and the Krylov-subspace dimension $m$. The values we have used are $\textrm{dt} = 0.02$ and $m=15$.

\section{Dipole moment analysis}
We show in \refFig{dipole} the results of exact diagonalization of a half-filled fermionic system where each point represents an
eigenstate in the space of energy and dipole-moment. As expected, in a given energy window and for large field (\refFig{dipole}a) the many body wave functions have well-defined dipole moment.
For a weak field however (\refFig{dipole}b), this is not the case. The eigenstates in a given energy window span a range of dipole moments. Around the critical field (\refFig{dipole}c,d), while the eigenstates in a given energy have a finite spread in the dipole moment, the different sectors become distinct
and the integer part of the dipole moment behaves as a conserved quantity.


\bibliographystyle{natbib}

\end{document}